\documentclass[%
amsmath,
amssymb,
aps,
prfluids
]{revtex4-2}

\usepackage{graphicx}
\usepackage{dcolumn}
\usepackage{bm}

\begin{document}


\title{Determination of the Young's angle using static friction in capillary bridges}

\author{Jong-In Yang}
 \email{jiyang@hanyang.ac.kr}
\author{Jooyoo Hong}%
 \email{jhong@hanyang.ac.kr}
\affiliation{%
Department of Applied Physics, Hanyang University ERICA, 55 Hanyangdeahak-ro, Sangnok-gu, Ansan, Gyeonggi-do, 15588, Republic of Korea
}%

\date{\today}

\begin{abstract}
Recently contact angle hysteresis in two-dimensional droplets lying on a solid surface has been studied extensively in terms of static friction due to pinning forces at contact points. Here we propose a method to determine the coefficient of static friction using two-dimensional horizontal capillary bridges. This method requires only the measurement of capillary force and separation of plates, dispensing with the need for direct measurement of critical contact angles which is notoriously difficult. Based on this determination of friction coefficient, it is possible to determine the Young's angle from its relation to critical contact angles (advancing or receding). The Young's angle determined with our method is different either from the value estimated by Adam and Jessop a hundred years ago or the value argued by Drelich recently, though it is much closer to Adam and Jessop's numerically. The relation between energy and capillary force shows a capillary bridge behaves like a spring. Solving the Young-Laplace's equation, we can also locate the precise positions of neck or bulge and identify the exact moment when a pinch-off occurs.
\end{abstract}

\maketitle


\section{Introduction}
Contact angle hysteresis of a droplet on a solid surface has been well-known for over a century \cite{Adam_Jessop_1925, Adam_1941, Good_1952, Drelich_2020, Butt_2022}. The cause of this phenomenon was rightly pointed to be the static friction between the liquid drop and the solid surface by Adam and Jessop \cite{Adam_Jessop_1925}. However, the discussions have remained qualitative for a long while, lacking detailed investigations of its physical origin or precise quantitative analyses \cite{Gao_2018, Drelich_2020, Butt_2022, McHale_2022, Zhang_2023}. 

In our recent work \cite{Yang_arXiv_2024}, we have attempted and succeeded in explaining quantitatively the phenomenon of contact angle hysteresis using static friction at the solid-liquid contact point of a two-dimensional droplet. We defined the coefficient of static friction and provided a systematic quantitative analysis of its mechanical implications. Though this method is useful in understanding the phenomenon, it presents several practical difficulties: (1) A two-dimensional droplet is a mathematical idealization and difficult to implement experimentally \footnote{See, as an exception, a recent paper by Lv and Shi in which they devised an experiment for a two-dimensional droplet system: C. Lv and S. Shi, Wetting states of two-dimensional drops under gravity, Phys. Rev. E \textbf{98}, 042802 (2018)}. (2) The contact angles of droplets must be measured directly in experiments to determine the coefficient of static friction, which is difficult to measure accurately \cite{Drelich_1994, Extrand_1995, Krasovitski_2005, Pozzato_2006, Pierce_2008, Ryan_2008, Kalantarian_2009, Krumpfera_2010, Xu_2010, Rodriguez-Valverde_2010, Ruiz-Cabello_2011, Ruiz-Cabello_2011_2, Srinivasan_2011, Behroozi_2012, Prydatko_2018, Chen_2018, Ravazzoli_2019, Zhang_2021, Beitollahpoor_2022, Klauser_2022, Wood_2023, Shumaly_2023, Johnson_2024, Chen_2024, Ruiz-Cabello_2014, Drelich_2019} and thus has some subtle and controversial issues due to hysteresis \cite{Drelich_2013, Ruiz-Cabello_2014, Drelich_2019, Huang_2020, Kleist-Retzow_2020, Marmur_2022, Kadyrov_2023}. 

In the present work we will consider two-dimensional horizontal capillary bridges instead of two-dimensional droplets, and propose a method to determine the coefficient of static friction without directly measuring contact angles. As a direct consequence this enables a new way of determining the Young's angle without ambiguity. Since there was a recent work by Drelich \cite{Drelich_2019} arguing that the traditional way of defining and measuring the Young's angle is mistaken, our work will shed a new light on this issue. 

In contrast to contact angles which are difficult to measure precisely, the capillary force and the height of capillary bridges can be measured with high precision \cite{Willett_2000, Rabinovich_2005, Cheneler_2008, Souza_2008, Souza_2008_2, Nagy_2019, Nagy_2022, Daniel_2023}. By measuring these quantities accurately the contact angles of the capillary bridge can be deduced precisely. The coefficient of static friction can then be calculated from the measured values of critical contact angles. In turn the Young's angle is to be determined. It is interesting to note that in the absence of gravity we only need the separation between two plates to obtain the contact angles.

Capillary bridges, like droplets, have long been studied theoretically, experimentally, and through numerical works, mostly in three dimensions. There is no shortage of works on these topics in the literature \cite{Kralchevsky_2001, Langbein_2002, Mastrangeli_2015}. For example, there are studies investigating three-dimensional capillary bridges between two plates \cite{Fortes_1982, Boucher_1982, Meseguer_1985, Rive_1986, Lowry_1995, Concus_1998, Tadrist_2019, Nguyen_2020, Chan_2021, Carter_1988, Pratt_2021, Chen_2013, Akbari_2016, Odunsi_2023}; between a sphere and a plate \cite{Honschoten_2010, Nguyen_2019, Orr_1975}; between a cylinder and a plate \cite{Reyssat_2015}; between two cylinders \cite{Cooray_2016}; and between two spheres \cite{Lian_2016, Kruyt_2017, Lian_1993}, etc. There are also some studies on two-dimensional capillary bridges between two parallel plates \cite{Broesch_2012, Teixeira_2019, Wang_2019, Pang_2021}. As in the case of droplets the contact angle hysteresis should also exist for all of these capillary bridges \cite{Chen_2013, Akbari_2016, Odunsi_2023}. 

In the past theoretical studies on two-dimensional horizontal capillary bridges, they rely on numerical solutions to the Young-Laplace's equation and treat the contact angle as a free parameter. And the relationship between the capillary force and the energy of the capillary bridge has not been quantitatively established. As in our previous work we derive the Young's equation modified due to contact angle hysteresis and provide exact analytical solutions to the Young-Laplace's equation for two-dimensional horizontal capillary bridges. Furthermore, we derive and interpret the relationship between the energy of the capillary bridge and the capillary force, as only qualitatively mentioned by Teixeira et al. \cite{Teixeira_2019}.

\begin{figure}[t!]
    \centerline{\includegraphics[width=0.6\textwidth]{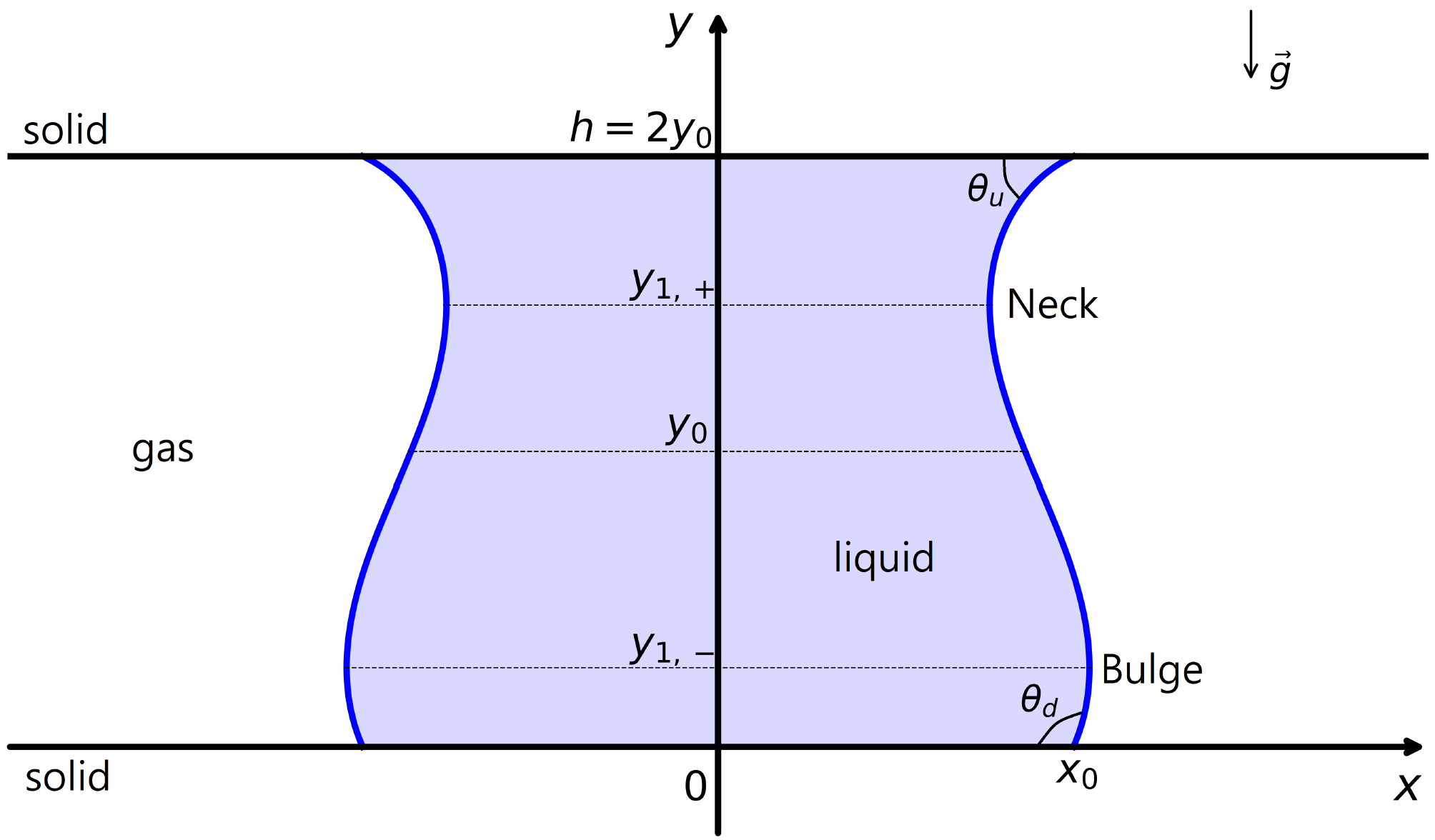}}
    \caption{A generic shape of a two-dimensional horizontal capillary bridge with height (i.e. separation) $h = 2 y_{0}$ and cross-sectional area $A$. Here, $y_{1, \pm}$ are positions of 'neck' or 'bulge'.}
    \label{Fig_1}
\end{figure}

As a set-up, let us consider a two-dimensional capillary bridge between two horizontal planar solid plates. The lower plate is assumed to be fixed while the upper plate is supposedly very light and can change its elevation freely. As we see in Fig.\ref{Fig_1} the shapes of capillary bridges differ by several factors: total cross-sectional area, total contact length, contact angles, separation (or height) of capillary bridge, and so on. The conditions and the equations that govern the profile of a static capillary bridge of cross-sectional area $A$ are given as follows: First, the profile of a capillary bridge is related to its area $A$ by
\begin{equation} \label{eqn: constraint_conditions}
    A = 2 \int_{0}^{2 y_{0}} x \left(y\right) \,dy
\end{equation}
where $x \left(y\right)$ is the profile function for $x \geq 0$. (It is enough to consider only the right half due to the left-right symmetry.) Second, since we will assume the contact points are fixed due to static friction, we have
\begin{equation*}
    x \left(0\right) = x_{0}, \qquad x \left(2 y_{0}\right) = x_{0}\ .
\end{equation*}
Then the total contact length $2 x_{0}$ is a fixed parameter. And last, the profile of the capillary bridge will be determined by minimizing its energy (\textit{energy/length}, in fact) while satisfying the constraints in the above.
\begin{align} \label{eqn: mathcal_E}
    \begin{split}
        \mathcal{E} &= \gamma \int_{0}^{2 y_{0}} \sqrt{1 + \left\{x' \left(y\right)\right\}^{2}} \,dy - \Delta \gamma x \left(2 y_{0}\right) - \Delta \gamma x \left(0\right) + \rho g \int_{0}^{2 y_{0}} y x \left(y\right) \,dy\\
                    & \quad - \beta \left[\int_{0}^{2 y_{0}} x \left(y\right) \,dy - \frac{A}{2}\right] - \kappa_{u} \left[x \left(2 y_{0}\right) - x_{0}\right] + \kappa_{d} \left[x \left(0\right) - x_{0}\right]
    \end{split}
\end{align}
where $\Delta \gamma \equiv \gamma^{sg} - \gamma^{sl} = \gamma \cos \theta_{Y}$ ($\gamma^{sg}$, $\gamma^{sl}$, and $\gamma$ are the surface tension between solid-gas, solid-liquid, and liquid-gas, respectively), and the last three terms are needed to give the constraint conditions ($\beta$, $\kappa_{u}$, and $\kappa_{d}$ are the Lagrange multipliers). The modified Young's equation and the profile equation (the Young-Laplace's equation) of a capillary bridge can be obtained by the energy minimization (see Appendix A, B).

Let us introduce dimensionless variables as $X \equiv x/x_{0}$, $Y \equiv y/x_{0}$, $H \equiv h/x_{0} = 2 Y_{0}$, and $A^{*} \equiv A/x_{0}^{2}$. And the Bond number is defined as $B \equiv \rho g x_{0}^{2} / \gamma$ where $\rho$ and $g$ are the density of liquid and the acceleration of gravity, respectively.

In a capillary bridge without gravity, the system has the up-down symmetry about $y = y_{0}$, so the upper and down contact angles are the same (i.e. $\theta_{u} = \theta_{d} \equiv \theta_{C}$). The equilibrium contact angle ($\theta_{eq}$) and the Young's angle ($\theta_{Y}$) are generally not the same due to hysteresis. When gravity is turned on, the up-down symmetry will be broken, and thus the upper contact angle decreases (similar to a pendent droplet), while the down contact angle increases (similar to a sessile droplet) (i.e. $\theta_{u} \leq \theta_{C} \leq \theta_{d}$).

\section{Energy of capillary bridges and Capillary forces}
The energy (per unit length) of two-dimensional horizontal capillary bridges under gravity was introduced in Eq.(\ref{eqn: mathcal_E}). Redefining the energy as $E \equiv \mathcal{E} / \left(\gamma x_{0}\right) + 2 \cos \theta_{Y}$, dimensionless and independent of the Young's angle $\theta_{Y}$,
\begin{equation} \label{eqn: E}
    E = \int_{0}^{2 Y_{0}} \sqrt{1 - \left\{f \left(Y\right)\right\}^{2}} \,dY + \frac{4 Y_{0} - A^{*}}{4 Y_{0}} \left(\cos \theta_{u} + \cos \theta_{d}\right) + \frac{1}{2} A^{*} B Y_{0}
\end{equation}
where $f \left(Y\right)=(B/2)(Y-\bar{Y})^2-\bar{f}$, $\bar{Y} = Y_{0} - (\cos \theta_{u} + \cos \theta_{d})/(2 B Y_{0})$, and $\bar{f} = B\bar{Y}^{2}/2 + \cos \theta_{d}$. The first term on the right side can be integrated and expressed in elliptic functions \cite{Gradshtein_2014}. Integration must be done separately depending on the range of $\bar{Y}$: (1) $0 \leq \bar{Y} \leq 2 Y_{0}$, (2) $\bar{Y} > 2 Y_{0}$, and (3) $\bar{Y} < 0$. For the last two ranges of $\bar{Y}$, $\bar{f}$ must be divided into the two cases of either $\left|\bar{f}\right| \leq 1$ or $\bar{f} > 1$. 

When both the gravity and the capillary force exist and compete with each other to shape the bridges, the above energy expression looks too complicated to deal with. Let us first single out the effect of pure capillary force (the case of no gravity). The expression for the energy reduces to
\begin{equation*}
    E_{C} = \int_{0}^{2 Y_{0}} \sqrt{1 - \left\{f_{C} \left(Y\right)\right\}^{2}} \,dY + \frac{4 Y_{0} - A^{*}}{2 Y_{0}} \cos \theta_{C} = \frac{\pi - 2 \theta_{C}}{\cos \theta_{C}} Y_{0}
\end{equation*}
where $f_{C} \left(Y\right) = (Y/{Y_{0}} - 1) \cos \theta_{C}$. We find $E_{C}$ is always positive. 

The profile of a two-dimensional liquid bridge is determined as the solution to the following force-balance equation in the horizontal direction.
\begin{equation*}
    p(y)-p_0=-\gamma \frac{d}{dy}\cos\theta(y)
\end{equation*}
where $p(y)=p_u+\rho g (2y_0-y)$, $p_0$ the gas pressure, and $p_u$ the liquid pressure at the upper plate. Integrating both sides with respect to $y$ from $0$ to $2y_0$ and using the definitions of contact angles $\theta_u=\theta(2y_0)$ and $\theta_d=\pi-\theta(0)$, we get 
\begin{equation*}
    p_{u} - p_{0} = - \rho g y_{0} - \gamma \frac{\cos \theta_{u} + \cos \theta_{d}}{2 y_{0}} \ .
\end{equation*}
Likewise, the force-balance in the vertical direction results in the following relation
\begin{equation} \label{eqn: normal_eq_for_bridge}
    2 \gamma \left(\sin \theta_{u} - \sin \theta_{d}\right) = \rho g \left(A - 4 x_{0} y_{0}\right) \ .
\end{equation}
This becomes the constraint condition for the cross-sectional area $A$. The dimensionless form of the above constraint condition is given in Appendix A (Eq.(\ref{eqn: area_constraint_with_gravity})).

As emphasized in our recent work \cite{Yang_arXiv_2024}, the contact angle at each plate is not equal to the Young's angle due to hysteresis. At each contact point we have the force-balance equations in the horizontal direction
\begin{equation} \label{eqn: kappa_ud}
    \kappa_{u} = \gamma \left(\cos \theta_{u} - \cos \theta_{Y}\right), \qquad \kappa_{d} = \gamma \left(\cos \theta_{Y} - \cos \theta_{d}\right)
\end{equation}
where $\kappa_u$ and $\kappa_d$ are the static frictions for the top and the bottom plate, respectively. These are the constraint forces needed to hold contact points at their given positions. (See Eq.(\ref{eqn: contact_point_constraint_with_gravity}) for more detailed explanations.) 

\begin{figure}[t!]
    \includegraphics[width=0.95\textwidth]{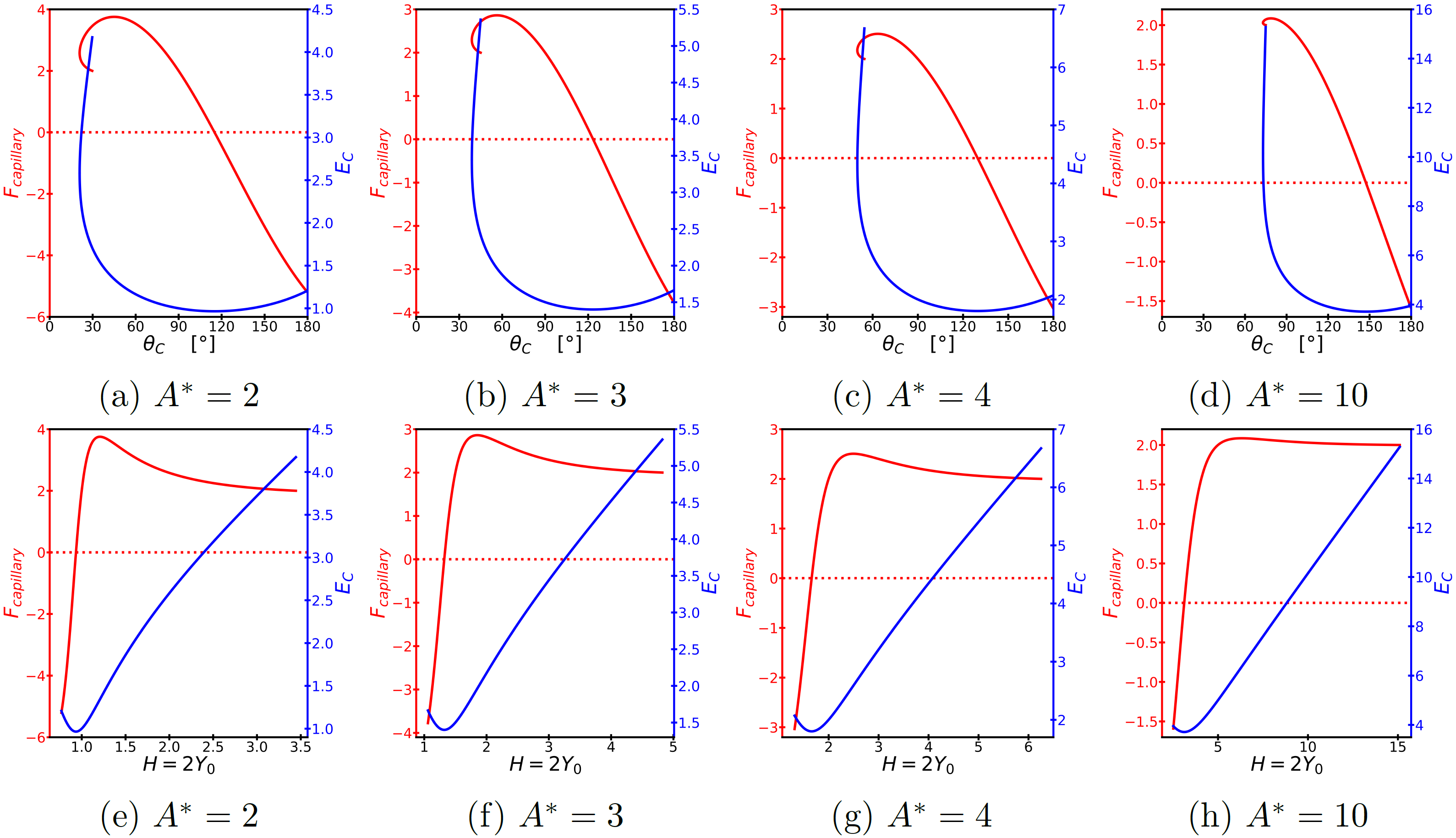}
    \caption{(a)-(d): $F_{\text{capillary}}$ (red), $E_{C} $ (blue) vs. $\theta_{C}$ and (e)-(h): $F_{\text{capillary}}$ (red), $E_{C}$ (blue) vs. $H = 2 Y_{0}$.}
\label{Fig_2}
\end{figure}

The vertical force component per length acting on the upper plate is, making it dimensionless by dividing by $\gamma$,
\begin{align} \label{eqn: F_capillary}
    \begin{split}
        F_{\text{capillary}} &= 2 \left(\sin \theta_{u} - \frac{ x_{0}}{\gamma} \left(p_{u} - p_{0}\right)\right)\\
                             &= 2 \sin \theta_{u} + 2 B Y_{0} + \frac{\cos \theta_{u} + \cos \theta_{d}}{Y_{0}}
    \end{split}
\end{align}
where $F_{\text{capillary}}$ is to be called the capillary force which is opposite to the force that should be applied externally to keep the bridge static. In this equation $2 \sin \theta_{u}$ is the upward pinning force due to the liquid-gas surface tension, and the next term is the force due to the pressure difference on each side of the upper plate. The force exerted by the capillary bridge on the upper plate should be defined as the opposite of $F_{\text{capillary}}$, i.e. $\mathcal{F} \equiv - F_{\text{capillary}}$. It is downward when $\mathcal{F} < 0$ (or $F_{\text{capillary}} > 0$), and is upward when $\mathcal{F} > 0$ (or $F_{\text{capillary}} < 0$). If $\mathcal{F} = F_{\text{capillary}} = 0$, the upper plate does not move even without being held externally. Without gravity ($B = 0$) Eq.(\ref{eqn: F_capillary}) becomes
\begin{equation} \label{eqn: F_capillary_without_gravity}
    F_{\text{capillary}} = 2 \left(\sin \theta_{C} + \frac{ \cos \theta_{C}}{Y_{0}}\right) \ .
\end{equation}

In Eq.(\ref{eqn: E}), $E$ is the expression for the energy in the region $X \geq 0$, hence the total energy is $E_{\text{total}} = 2 E$ due to the left-right symmetry. By taking the derivative of $E_{\text{total}}$ with respect to $H = 2 Y_{0}$, we get
\begin{equation*}
    \frac{\partial E_{\text{total}}}{\partial H} = \frac{\partial E}{\partial Y_{0}} = 2 \sin \theta_{u} + 2 B Y_{0} + \frac{\cos \theta_{u} + \cos \theta_{d}}{Y_{0}} \ .
\end{equation*}
Therefore, the relation between the energy of capillary bridge and the force it exerts on the upper plate is given by
\begin{equation*}
    \mathcal{F} = - \frac{\partial E_{\text{total}}}{\partial H} \ .
\end{equation*}
Or equivalently, 
\begin{equation*}
    F_{\text{capillary}} = +\frac{\partial E_{\text{total}}}{\partial H} \ . 
\end{equation*}

\begin{figure}[t!]
    \includegraphics[width=0.95\textwidth]{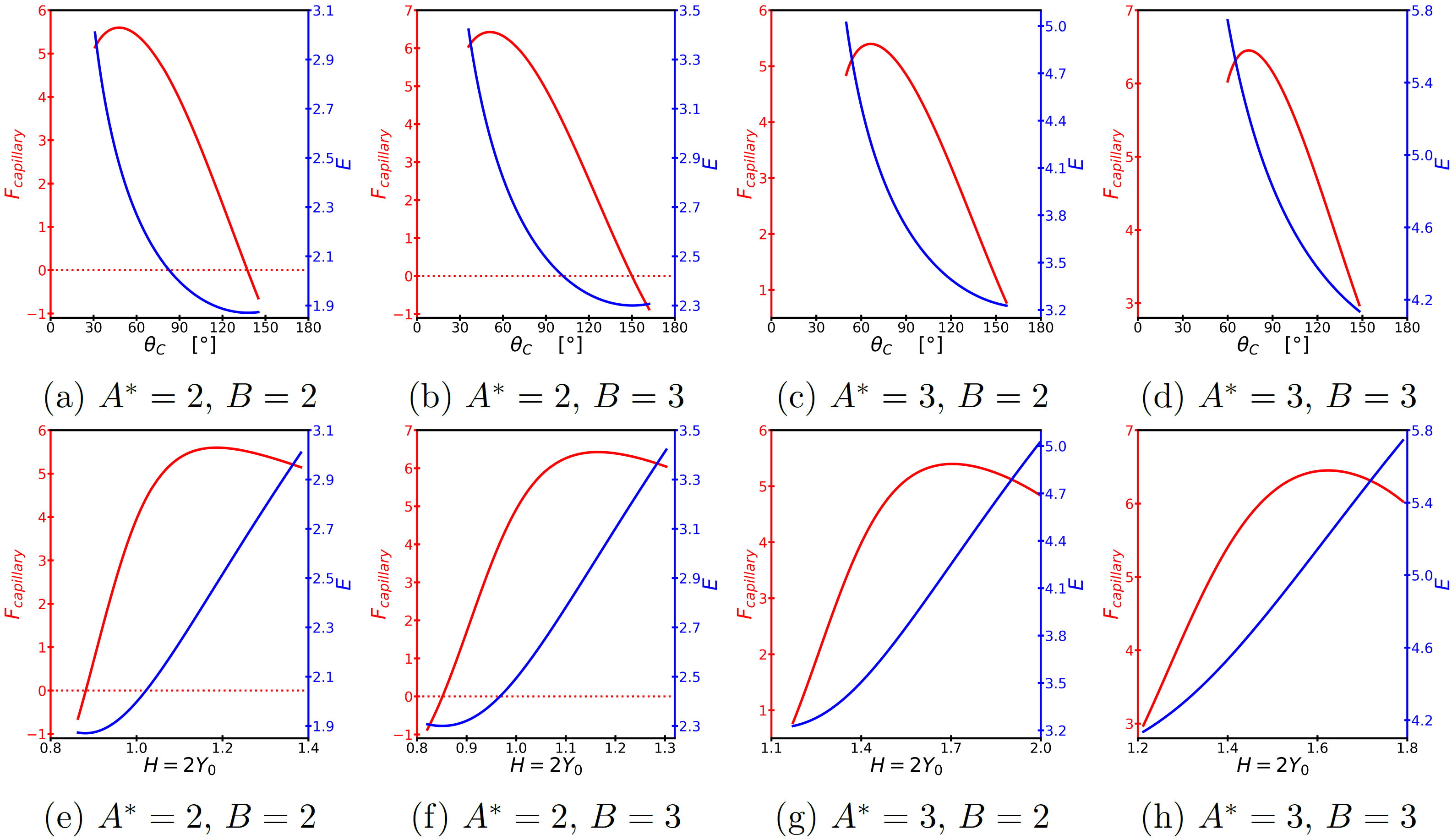}
\caption{(a)-(d): $F_{\text{capillary}}$ (red), $E $ (blue) vs. $\theta_{C}$ and (e)-(h): $F_{\text{capillary}}$ (red), $E$ (blue) vs. $H = 2 Y_{0}$. 
         Note that the equilibrium point does not appear for $A^{*} = 3$.}
\label{Fig_3}
\end{figure}

We find this equilibrium is stable since the second derivative of $E$ is always positive
\begin{equation*}
    \frac{\partial^{2} E}{\partial Y_{0}^{2}} = \frac{\partial F_{\text{capillary}}}{\partial Y_{0}} = - \frac{\partial \mathcal{F}}{\partial Y_{0}} \quad > 0 \ .
\end{equation*}
Therefore, the capillary bridges behave like springs \cite{Kusumaatmaja_2010, Sariola_2019}. We plotted $E_{C}$, $E$, and $F_{\text{capillary}}$ for some values of $A^{*}$ and $B$ in Figs.\ref{Fig_2} and \ref{Fig_3}.

For a small vertical displacement $\Delta H$ (or equivalently, small change of contact angles) from the equilibrium point the force exerted by the capillary bridge on the upper plate can be approximated as
\begin{equation*}
    \mathcal{F} \cong - \textrm{k}_{1} \Delta H - \textrm{k}_{2} \left(\Delta H\right)^{2}
\end{equation*}
where $\textrm{k}_{1}$ and $\textrm{k}_{2}$ are the effective linear and quadratic spring constants of the capillary bridge. These spring constants depend on the equilibrium values of $\theta_u$ and $\theta_d$ in a very complicated way. However, in the absence of gravity where there is only a single equilibrium angle $\theta_{eq}$ due to the up-down symmetry we can obtain simple expressions of $\textrm{k}_{1}$ and $\textrm{k}_{2}$.
\begin{equation*}
    \textrm{k}_{1} = -2 \sin \theta_{eq} \tan \theta_{eq} \qquad \text{and} \qquad \textrm{k}_{2} = -2 \sin \theta_{eq} \tan^{2} \theta_{eq}
\end{equation*}
These constants are different from the results of Kusumaatmaja and Lipowsky \cite{Kusumaatmaja_2010} and Sariola \cite{Sariola_2019} where they have dealt with axisymmetric 3-dimensional bridges.

In the absence of gravity, Eq.(\ref{eqn: F_capillary_without_gravity}) shows that the capillary force can only be zero for $90^{\circ} < \theta_{C} \leq 180^{\circ}$. Using the constraint condition for the area (Eq.(\ref{eqn: area_constraint_without_gravity})) and $Y_{eq} = - \cot \theta_{eq}$ ($\theta_{eq} > 90^{\circ}$), we can also obtain the relation between the equilibrium contact angle ($\theta_{eq}$) and the area:
\begin{equation*}
    A^{*} \sin^{2} \theta_{eq}+ \sin \left(2 \theta_{eq}\right) - 2 \theta_{eq} + \pi = 0 \ .
\end{equation*}
If we include gravity in our discussion (i.e. Eq.(\ref{eqn: F_capillary})), the corresponding relations become too complicated to be useful, so we refrain from writing them down. However, we find that there may not exist an equilibrium configuration for some ranges of area and the Bond number. The necessary condition for the equilibrium to exist turns out to be
\begin{equation*}
    B H_{\text{min}}^{4} - 2 A^{*} B H_{\text{min}}^{3} + \left(A^{*} B + 4 B + 4\right) H_{\text{min}}^{2} - 4 A^{*} \left(1 + B\right) H_{\text{min}} + A^{*2} B \geq 0
\end{equation*}
where $H_{\text{min}}$ is the minimum separation (i.e. when $\theta_{d} = 180^{\circ}$).

\section{Determination of the coefficient of static friction and the Young's angle}
\begin{figure}[t!]
    \includegraphics[width=0.95\textwidth]{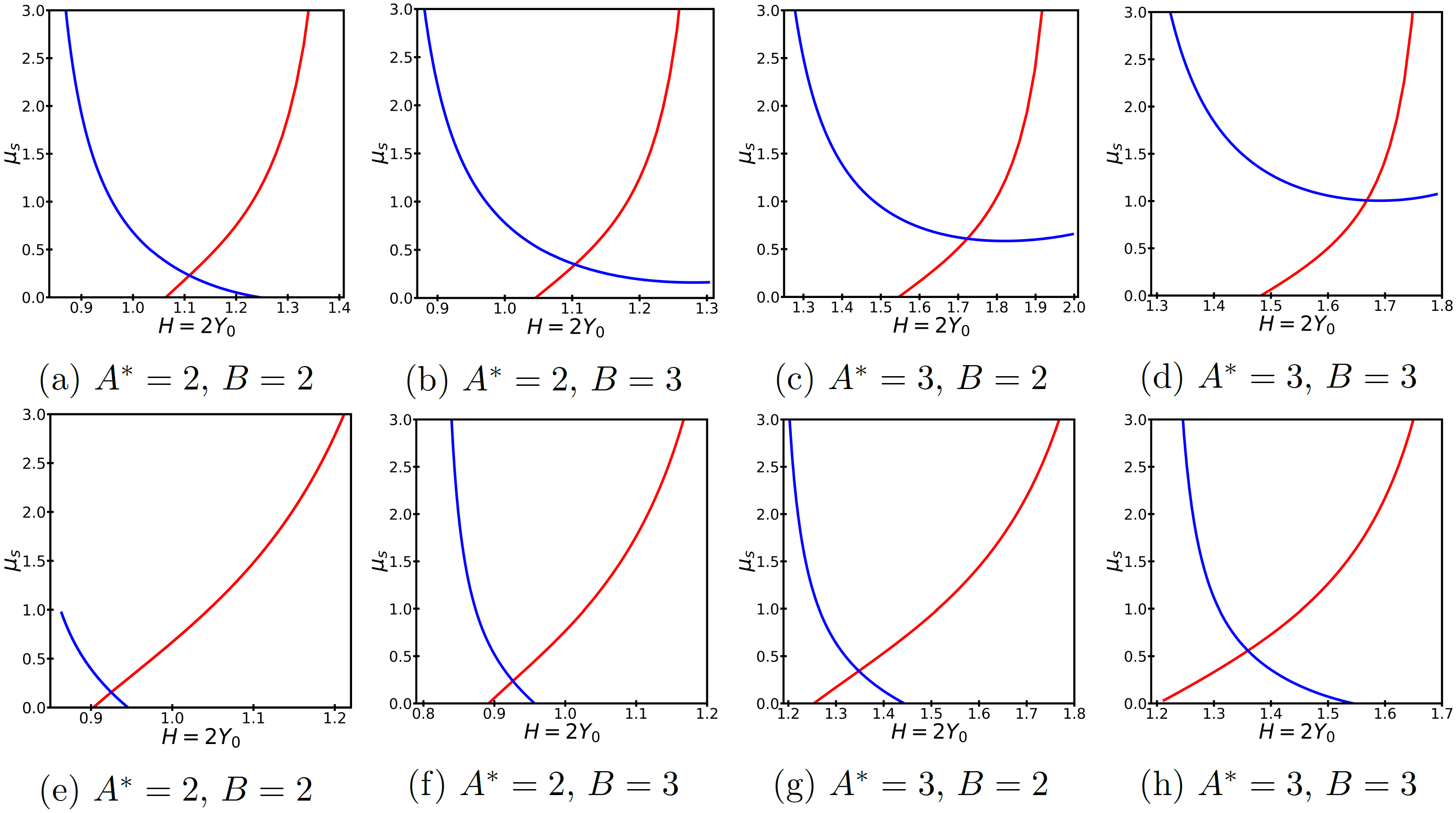}
    \caption{The coefficients of static friction $\mu_{s}$ for given conditions when $B > 0$.
             (a)-(d): $\theta_{Y} = 60^{\circ}$ and (e)-(h): $\theta_{Y} = 120^{\circ}$.
             Blue lines are for $\theta_{a} = \theta_{d}$, $\theta_{u} > \theta_{r}$, and red lines are for $\theta_{r} = \theta_{u}$, $\theta_{d} < \theta_{a}$ in Eq.(\ref{eqn: mu_1}), respectively.}
\label{Fig_4}
\end{figure}

As we have argued in our previous work on two-dimensional droplets \cite{Yang_arXiv_2024}, the coefficient of static friction is defined from the critical angles (advancing angle $\theta_{a}$ and receding angle $\theta_{r}$) as follows:
\begin{equation} \label{eqn: mu_1}
    \mu_{s} =
    \begin{cases} 
        \dfrac{\cos \theta_{Y} - \cos \theta_{a}}{\sin \theta_{a}} \qquad \qquad \qquad \left(\theta_{a} = \theta_{d}, \quad \theta_{u} > \theta_{r}\right) \\
        \\
        \dfrac{\cos \theta_{r} - \cos \theta_{Y}}{\sin \theta_{r}} \qquad \qquad \qquad \left(\theta_{r} = \theta_{u}, \quad \theta_{d} < \theta_{a}\right)\ .
    \end{cases}
\end{equation}
For capillary bridges it is obvious that $\theta_u \leq\theta_d$. If we push down the upper plate, both $\theta_u$ and $\theta_d$ increase and hence $\theta_d$ reaches the advancing angle $\theta_{a}$ first. We can determine $\mu_{s}$ using the first expression of Eq.(\ref{eqn: mu_1}). Or, if we pull up the upper plate, both $\theta_u$ and $ \theta_d$ decrease, hence $\theta_u$ reaches the receding angle $\theta_{r}$ first. The second expression of Eq.(\ref{eqn: mu_1}) then gives $\mu_{s}$. Fortunately, unlike the case of droplets we have two ways to determine the coefficient of static friction for bridge systems. We have plotted the graphs of $\mu_{s}$ vs. $H$ in Fig.\ref{Fig_4} for various values of $A^{*}$ and $B$.

It is well known that we can precisely measure the capillary force using the instruments like capillary bridge probes \cite{Nagy_2019, Nagy_2022}. Using the measured values of capillary force and gap we can calculate the contact angles, thus deducing the precise values of contact angles. The relevant relations are given in Eq.(\ref{eqn: normal_eq_for_bridge}) (or Eq.(\ref{eqn: area_constraint_with_gravity})) and Eq.(\ref{eqn: F_capillary}). This situation is in sharp contrast to the case of direct measurement of contact angles which is usually plagued by difficulty or inaccuracy. 

\begin{figure}[t!]
    \includegraphics[width=0.95\textwidth]{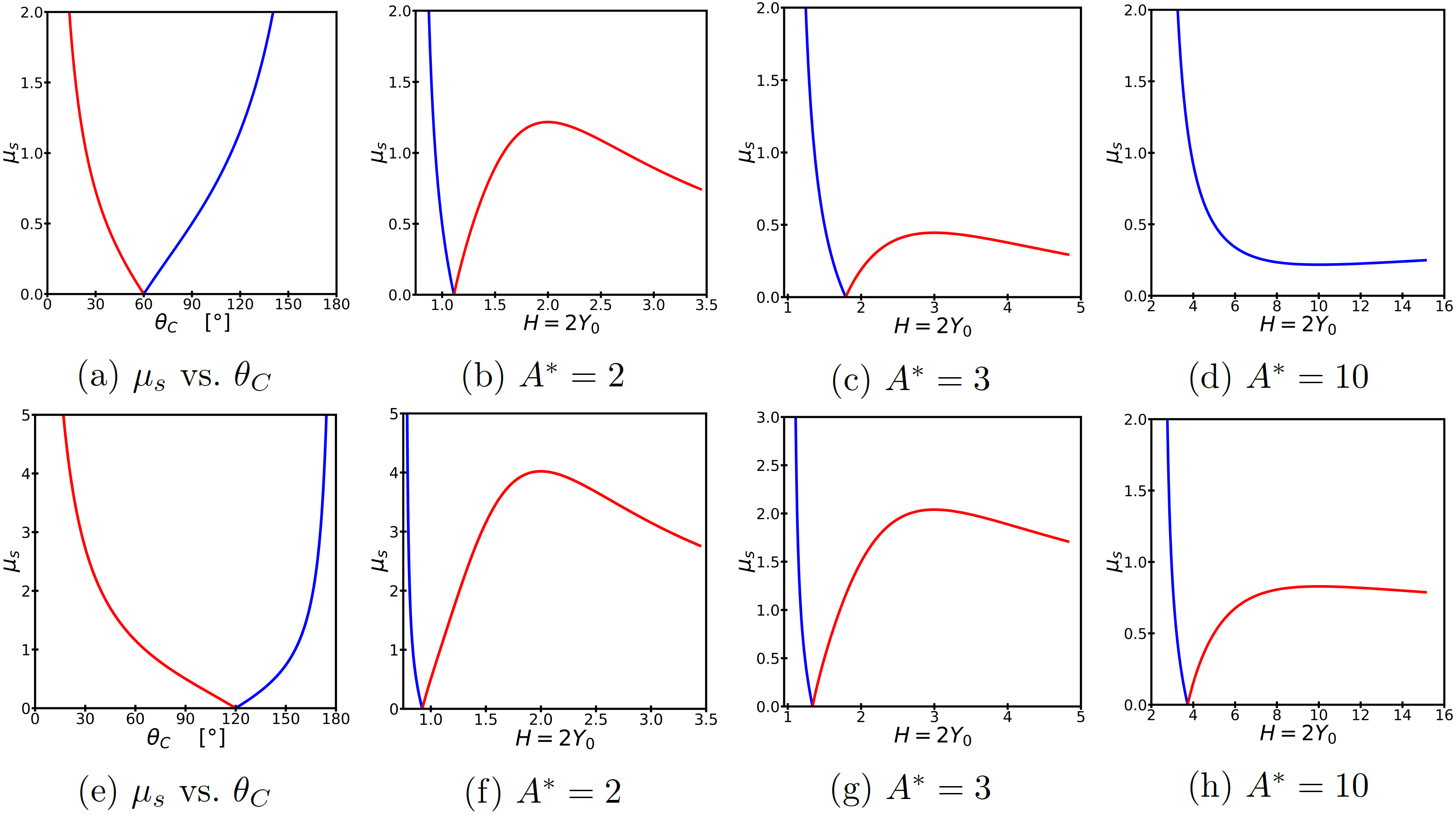}
    \caption{$\mu_{s}$ vs. $\theta_{C}$ or $\mu_{s}$ vs. $H$ when $B = 0$.
             (a)-(d): $\theta_{Y} = 60^{\circ}$ and (e)-(h): $\theta_{Y} = 120^{\circ}$.
             Blue lines are for $\theta_{C} = \theta_{a} > \theta_{Y}$ and red lines are for $\theta_{C} = \theta_{r} < \theta_{Y}$.
             The graphs in (a) and (e) ($\mu_{s}$ vs. $\theta_{C}$) are independent of $A^{*}$.
             For the graphs $\mu_{s}$ vs. $H$ the rightmost points correspond to the pinch-offs. 
             The minimum contact angle ($\theta_{C, \text{min}}$) depends on $A^{*}$ (See Eq.(\ref{eqn: theta_c_min}).). 
             For the graph in (d) ($\theta_{Y} = 60^{\circ}$ and $A^{*} = 10$), the receding contact angle is not drawn because $\theta_{C, \text{min}} \cong 73.01^{\circ}>\theta_{Y} = 60^{\circ}$.}
\label{Fig_5}
\end{figure}

If there is no gravity (i.e. $B = 0$), there is a further simplification. The contact angle can be obtained simply by measuring the height $H$; $\theta_{C}$ can be calculated from the height (Eq.(\ref{eqn: area_constraint_without_gravity})). This is summarized in Fig.\ref{Fig_5}. In this case we have both the left-right and the up-down symmetry, and the coefficient of static friction is described by the single formula
\begin{equation} \label{eqn: mu_2}
    \mu_{s} = \frac{\left|\cos \theta_{C} - \cos \theta_{Y}\right|}{\sin \theta_{C}} \qquad \qquad \left(\theta_{C} = \theta_{r} < \theta_{Y} \quad \text{or} \quad \theta_{C} = \theta_{a} > \theta_{Y}\right) \ .
\end{equation}

Since the coefficient of static friction should be unique for a given solid-liquid interface, the Young's contact angle can be determined by equating the two expressions in Eq.(\ref{eqn: mu_1}):
\begin{equation} \label{eqn: YoungAngle_1}
    \theta_{Y} = \cos^{-1} \left(\frac{\sin \left(\theta_{a} + \theta_{r}\right)}{\sin \theta_{a} + \sin \theta_{r}}\right) \ .
\end{equation}
In Eq.(\ref{eqn: kappa_ud}) we notice that the maximum static friction takes different values whether we push down the upper plate ($\theta_{a}$) or we pull up ($\theta_{r}$). This difference is caused by whether the contact points try to move toward solid-liquid interface ($\theta_{r}$) or toward solid-gas interface ($\theta_{a}$). 

However, in the previous studies \cite{Adam_Jessop_1925, Adam_1941, Good_1952, Drelich_2020}, it was simply assumed that the maximum static frictions at the advancing angle and at the receding angle are equal, which is, we think, unfounded. In the literature, the following relations are argued to hold
\begin{equation} \label{eqn: YoungAngle_2}
    \theta_{Y} = \cos^{-1} \left(\frac{\cos \theta_{a} + \cos \theta_{r}}{2}\right)\ .
\end{equation}
Or
\begin{equation*}
    \theta_{Y} \approx \frac{\theta_{a} + \theta_{r}}{2} \ .
\end{equation*}
From Eq.(\ref{eqn: YoungAngle_1}) the above relations are valid only when the critical angles $\theta_{a}$ and $\theta_{r}$ deviate slightly from $\theta_{C}$. 

In our previous work on droplets, we argued that the maximum static friction should also be proportional to $\sin \theta_{c}$ where $\theta_{c}$ is the critical contact angle. If the maximum static frictions at the advancing angle and at the receding angle are equal to each other, this implies $\theta_{a} + \theta_{r} = 180^{\circ}$. Putting this relation into the above equations quoted from the literature, we are led to the value of the Young's angle $\theta_{Y} = 90^{\circ}$, which is obviously wrong. So we conclude that instead of identifying the equilibrium contact angle or the averaged (or mean) contact angle as the Young's angle we have to use the true relation in Eq.(\ref{eqn: YoungAngle_1}). The Young's angle determined with our method is much closer to Adam and Jessop's (Eq.(\ref{eqn: YoungAngle_2})) than Drelich's \cite{Drelich_2019} numerically. 

Substituting Eq.(\ref{eqn: YoungAngle_1}) back into Eq.(\ref{eqn: mu_1}) or Eq.(\ref{eqn: mu_2}), the two expressions for the coefficient of static friction $\mu_{s}$ is merged into a single formula 
\begin{equation} \label{eqn: mu_3}
   \mu_{s} = \tan \left(\frac{\theta_{a} - \theta_{r}}{2}\right) \ .
\end{equation}
Hence $\mu_{s}$ can be determined solely from the critical contact angles without knowing the Young's angle $\theta_{Y}$. We note that $\theta_{a} = \theta_{r}$ if and only if $\mu_{s} = 0$.

\section{Conclusion}
We have obtained exact solutions to the profile equations for the two-dimensional horizontal capillary bridges between two plates with or without gravity. We studied the detailed shape of capillary bridges for a given set of parameters, hence locating the exact positions of neck or bulge, and finding the conditions for pinch-offs (See Appendix B, Fig.\ref{Fig_6} and Appendix C for mathematical details.).

Continuing from our previous work on contact angle hysteresis using static friction at the contact points, we have proposed a method to determine the coefficient of static friction and the Young's contact angle. This method utilizes the capacity to precisely measure the capillary force and the height of the capillary bridge, avoiding the direct measurements of the contact angles which are notoriously difficult and imprecise. From these accurately measured values of capillary force and height, it is possible to deduce the critical contact angles. Once we find the values of critical angles, using Eq.(\ref{eqn: YoungAngle_1}) and Eq.(\ref{eqn: mu_3}), the Young's angle and the coefficient of static friction are readily obtained.

Capillary bridge system is distinct from 2-dimensional droplets in that an equilibrium state can be broken even before the contact angle reaches its critical value, namely a pinch-off may occur while we increase the separation between the plates (see Fig.\ref{Fig_6} (g)). Instead of increasing the gap we may also control the area, the Bond number, or both to observe a pinch-off before any contact points move (see Fig.\ref{Fig_6} (m)).

The coefficient of static friction and the Young's contact angle, measured in this way, can be applied to other capillary phenomena including droplets, two-dimensional or three-dimensional, which would deepen our understanding of this fascinating area of research.

As a future study it would be interesting to investigate the cases of horizontally moving the upper plate instead of the vertical movement considered here. Since the left-right symmetry breaks down, the analysis would be much more complicated than the one given here, but we could have a good chance to better understand workings of static friction in capillary bridge systems. In addition, we want to point out that this method can also be applied to vertical capillary bridges or other kinds of capillary bridges of two or three dimensions.

\begin{acknowledgments}
One of the authors (J.-I. Y) would like to thank Profs. Young-Dae Jung and Bo Soo Kang for their support and encouragement throughout the course of his graduate years.
\end{acknowledgments}

\appendix
\section{Energy minimization and Force-balance equations}
The modified Young's equations, the Young-Laplace's equation, and the constraint forces ($\kappa_{u}$, $\kappa_{d}$, and $\beta$) can be derived by minimizing the energy of a generic capillary bridge in Eq.(\ref{eqn: mathcal_E}).

First, the modified Young's equation at each contact point is obtained as
\begin{equation} \label{eqn: contact_point_constraint_with_gravity}
    \Delta \gamma = \gamma \cos \theta_{Y} = \gamma \cos \theta_{u} - \kappa_{u}, \qquad \Delta \gamma = \gamma \cos \theta_{Y} = \gamma \cos \theta_{d} + \kappa_{d}\ .
\end{equation}
$\kappa_{u}$ and $\kappa_{d}$ are the constraint forces at contact points (Eq.(\ref{eqn: kappa_ud})). If we ignore the gravity ($B = 0$), $\theta_u=\theta_d\equiv \theta_{C}$ and the two equations merge to become
\begin{equation*}
    \Delta \gamma = \gamma \cos \theta_{Y} = \gamma \cos \theta_{C} - \kappa_{C}\ .
\end{equation*}
When the gap between the two plates are fixed, the contact points tend either to spread outward ($\theta_{C} > \theta_{Y}$) or to shrink inward ($\theta_{C} < \theta_{Y}$).

Second, the dimensionless form of the Young-Laplace's equation is derived as
\begin{equation} \label{eqn: YL_eq_1}
    - \frac{d}{dY} \left(\frac{X' \left(Y\right)}{\sqrt{1 + \left\{X' \left(Y\right)\right\}^{2}}}\right) + B Y - \beta^{*} = 0
\end{equation}
where $\beta^{*}$ is the dimensionless constraint force corresponding to the fixed area (see Eq.(\ref{eqn: beta})).

Lastly, by varying over the Lagrange multiplier $\beta$ we get the constraint condition for the fixed area in Eq.(\ref{eqn: constraint_conditions}) in terms of dimensionless parameters:
\begin{equation} \label{eqn: area_constraint_with_gravity}
    2 \left(\sin \theta_{u} - \sin \theta_{d}\right) = B \left(A^{*} - 4 Y_{0}\right) \ .
\end{equation}
This is the dimensionless version of the vertical force-balance equation for the capillary bridge in Eq.(\ref{eqn: normal_eq_for_bridge}). By integrating the Young-Laplace's equation for the entire interval $0\leq Y\leq 2Y_0$, we get 
\begin{equation} \label{eqn: beta}
    \beta^{*} \equiv \frac{\beta}{\gamma x_{0}} = B Y_{0} - \frac{\cos \theta_{u} + \cos \theta_{d}}{2 Y_{0}} \ .
\end{equation}
We can interpret $\beta$ as the pressure difference at the down plate (i.e. $\beta = p_{d} - p_{0}$ where $p_d$ is the liquid pressure at the down plate) or the total normal force per unit length acting on the capillary bridge by the down plate.

\section{Solving the Young-Laplace's equation}
We can solve the equations in Appendix A and draw the profiles of capillary bridges under given conditions.

\subsection{Without gravity ($B = 0$)}
When $B = 0$, Eq.(\ref{eqn: YL_eq_1}) and Eq.(\ref{eqn: beta}) reduce to
\begin{equation*}
    - \frac{d}{dY} \left(\frac{X_{C}' \left(Y\right)}{\sqrt{1 + \left\{X_{C}' \left(Y\right)\right\}^{2}}}\right) - \beta_{C}^{*} = 0, \qquad \beta_{C}^{*} = - \frac{\cos \theta_{C}}{Y_{0}}\ .
\end{equation*}
The circular profile, $X_{C} \left(Y\right)$, is obtained by integrating the above Young-Laplace's equation \cite{Delaunay_1841}.
\begin{equation*}
    X_{C} \left(Y\right) = 1 + Y_{0} \tan \theta_{C} - \frac{1}{\cos \theta_{C}} \sqrt{Y_{0}^{2} - \left(Y - Y_{0}\right)^{2} \cos^{2} \theta_{C}}
\end{equation*}
It is part of a circle with radius $\left|Y_{0} / \cos \theta_{C}\right|$ centered at $\left(X,Y\right)=\left(1 + Y_{0} \tan \theta_{C}, Y_{0}\right)$. Since $X'_{C} \left(Y_{0}\right) = 0$ and $X''_{C} \left(Y_{0}\right) = \cos \left(\theta_{C}\right) / Y_{0}$, it has either a neck when the contact angle is acute or a bulge when the contact angle is obtuse.

A simple geometrical relation between the area ($A^{*}$), the height ($H = 2 Y_{0}$), and the contact angle ($\theta_{C}$) exists:
\begin{equation} \label{eqn: area_constraint_without_gravity}
    \left[\pi - 2 \theta_{C} - \sin \left(2 \theta_{C}\right)\right] Y_{0}^{2} - 4 Y_{0} \cos^{2} \theta_{C} + A^{*} \cos^{2} \theta_{C} = 0 \ .
\end{equation}
This quadratic equation for $Y_0$ has two solutions
\begin{equation*}
    Y_{0, \pm} = \frac{2 \cos^{2} \theta_{C} \pm \sqrt{4 \cos^{4} \theta_{C} - A^{*} \left[\pi - 2 \theta_{C} - \sin \left(2 \theta_{C}\right)\right] \cos^{2} \theta_{C}}}{\pi - 2 \theta_{C} - \sin \left(2 \theta_{C}\right)}\ .
\end{equation*}
These are the positions of a neck ($\theta_{C} < 90^{\circ}$) or a bulge ($\theta_{C} > 90^{\circ}$) (see the discussion in Appendix C and Fig.\ref{Fig_7}). 

The minimum value of $Y_{0}$ is attained at $\theta_{C} = 180^{\circ}$
\begin{equation*}
    Y_{0, \text{min}} = \frac{\sqrt{4 + \pi A^{*}} - 2}{\pi} \ .
\end{equation*}
Since there could be possibly a pinch-off of the bridge while we vary the height ($H=2Y_0$), the maximum value of $Y_0$ is divided into two separate cases
\begin{equation*}
    Y_{0, \text{max}} = 
    \begin{cases}
        \dfrac{2 - \sqrt{4 - \pi A^{*}}}{\pi}                                   & \qquad \qquad \left(\text{no pinch-off}, \quad \theta_{C} = 0^{\circ}                \quad \text{and} \quad 0 \leq A^{*} \leq 4 - \pi\right)\\
        \\
        \dfrac{\cos \theta_{C}}{1 - \sin \theta_{C}} \equiv Y_{0, \text{pinch}} & \qquad \qquad \left(\text{pinch-off},    \quad \theta_{C} = \theta_{C, \text{pinch}} \quad \text{and} \quad 4 - \pi < A^{*}\right)
    \end{cases}
\end{equation*}
where $\theta_{C, \text{pinch}}$ and $Y_{0, \text{pinch}}$ are the values of $\theta_{C}$ and $Y_0$ when pinch-off occurs. 

By solving the equation $\partial \theta_{C} / \partial Y_{0} = 0$, we can get the minimum contact angle ($\theta_{C, \text{min}}$) for a given area ($A^{*}\geq 4 / \pi$). If $0 < A^{*} \leq 4 / \pi$, $\theta_{C, \text{min}} = 0^{\circ}$. And if $A^{*}\geq 4 / \pi$, the following relation should hold
\begin{equation} \label{eqn: theta_c_min}
    4 \cos^{2} \theta_{C, \text{min}} - A^{*} \left[\pi - 2 \theta_{C, \text{min}} - \sin \left(2 \theta_{C, \text{min}}\right)\right] = 0 \ .
\end{equation}
In contrast to the case of droplets, the two different values of heights are possible for a single contact angle $\theta_{C}$ in the interval $\theta_{C, \text{min}} \leq \theta_{C} \leq \theta_{C, \text{pinch}}$ (see Fig.\ref{Fig_7} (a)).

\begin{figure}[t!]
    \includegraphics[width=1.0\textwidth]{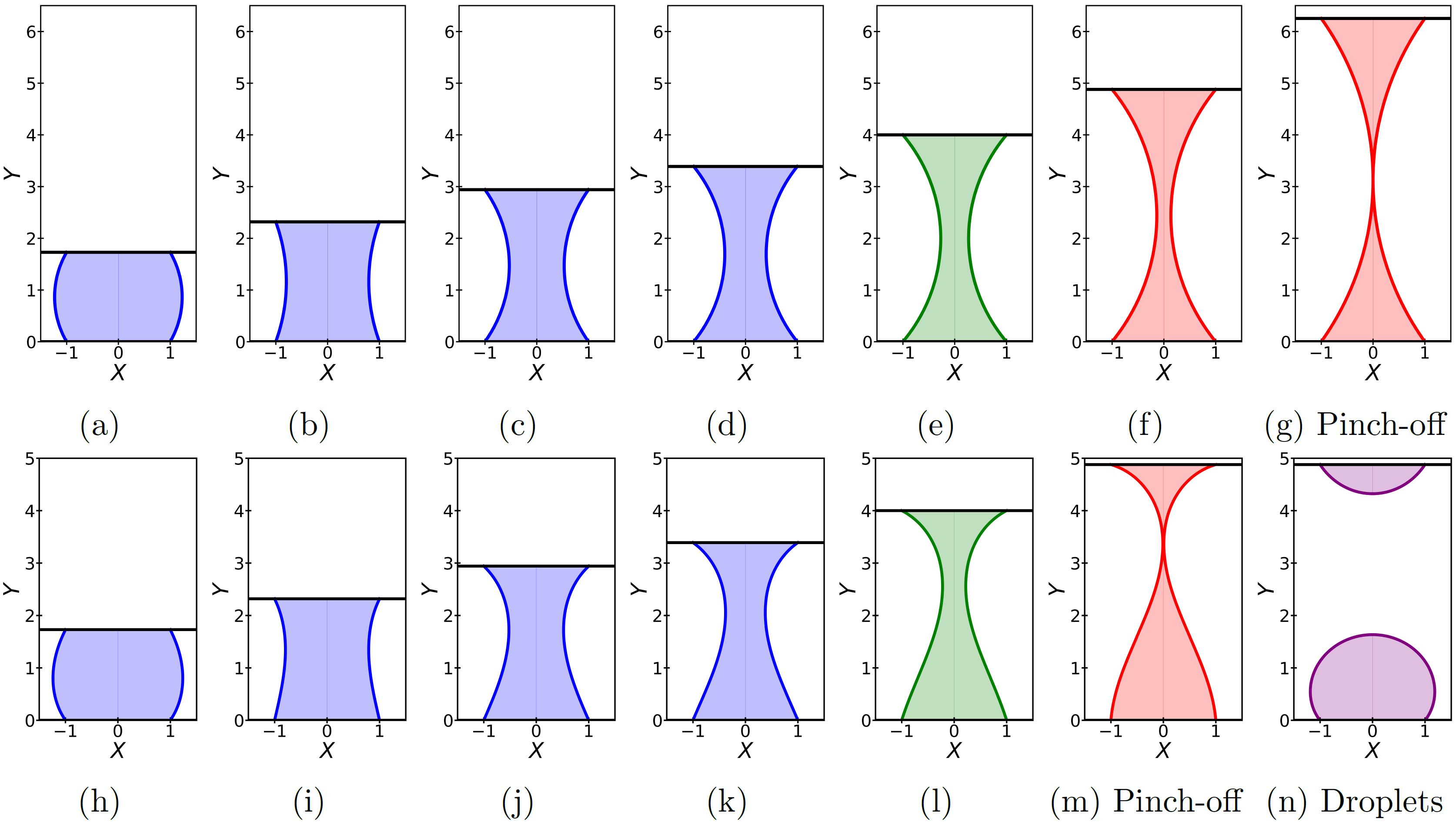}
    \caption{The shape of two-dimensional horizontal capillary bridges when $A^{*} = 4$ ($\theta_{C, \text{pinch}} \cong 54.52^{\circ}$, $\theta_{C, \text{min}} \cong 49.85^{\circ}$). The exact solutions are given in Appendix B.
            Blue, green, and red bridges correspond to $Y_{0,-}$, $Y_{0} = A^{*} / 2$, and $Y_{0,+}$, respectively.
            (a)-(g) are for $B = 0$. From the left to the right, $\theta_{C} = 120^{\circ}$, $70^{\circ}$, $54.52^{\circ}$, $51^{\circ}$, $\theta_{C, \text{min}}$, $51^{\circ}$, and $\theta_{C, \text{pinch}}$.
            (h)-(n) are for $B = 0.25$. Each of these profiles are obtained by turning on the gravity ($B=0.25$) from (a)-(f), respectively. From the left to the right, $\left(\theta_{u}, \theta_{d}\right) \cong $ $\left(116.83^{\circ}, 124.48^{\circ}\right)$, $\left(63.54^{\circ}, 76.92^{\circ}\right)$, $\left(42.95^{\circ}, 66.34^{\circ}\right)$, $\left(35.01^{\circ}, 66.99^{\circ}\right)$, $\left(27.01^{\circ}, 72.45^{\circ}\right)$, $\left(16.11^{\circ}, 85.18^{\circ}\right)$, and $\left(\theta_{p} \cong 56.5^{\circ}, \theta_{s} \cong 127.9^{\circ}\right)$.
            In the figure (n), $\theta_{p}$ and $\theta_{s}$ are the pendent angle and the sessile angle, respectively, and we used the results from our previous work \cite{Yang_arXiv_2024} where $A_{\text{pendent}}^{*} \cong 0.772$ and $A_{\text{sessile}}^{*} \cong 3.22$.
            }
\label{Fig_6}
\end{figure}

\subsection{With gravity ($B > 0$)}
Substituting Eq.(\ref{eqn: beta}) into Eq.(\ref{eqn: YL_eq_1}) and integrating both sides, we get
\begin{equation} \label{eqn: YL_eq_2}
    X' \left(Y\right) = \frac{f \left(Y\right)}{\sqrt{1 - \left\{f \left(Y\right)\right\}^{2}}}
\end{equation}
where
\begin{equation} \label{eqn: f(Y)}
    f \left(Y\right) = \frac{B}{2} \left(Y - \bar{Y}\right)^{2} - \bar{f}, \qquad \bar{Y} = Y_{0} - \frac{\cos \theta_{u} + \cos \theta_{d}}{2 B Y_{0}}, \qquad \bar{f} = \frac{B}{2} \bar{Y}^{2} + \cos \theta_{d} \ .
\end{equation}
The exact profiles are obtained by integrating the Young-Laplace's equation in Eq.(\ref{eqn: YL_eq_2}). The solutions can be written in terms of the elliptic functions \cite{Dwight_1961}. We divided the domain of $\bar{Y}$ into three intervals:

\noindent \underline{(1) $0 \leq \bar{Y} \leq 2 Y_{0}$}
\begin{equation*}
    X \left(Y\right) = 
    \begin{cases}
        1 - \dfrac{1}{\sqrt{B}} \left[2 \left\{E\left(\phi_{0}, k\right) - E\left(\phi, k\right)\right\} - \left\{F\left(\phi_{0}, k\right) - F\left(\phi, k\right)\right\}\right] & \qquad \qquad (0 \leq Y \leq \bar{Y})\\
        \\
        1 + \dfrac{1}{\sqrt{B}} \left[2 \left\{E\left(\psi_{1}, k\right) - E\left(\psi, k\right)\right\} - \left\{F\left(\psi_{1}, k\right) - F\left(\psi, k\right)\right\}\right] & \qquad \qquad (\bar{Y} \leq Y \leq 2 Y_{0})
    \end{cases}
\end{equation*}
The continuity of $X \left(Y\right)$ at $Y = \bar{Y}$ requires
\begin{equation*}
    2 \left[2 E\left(\frac{\pi}{2}, k\right) - E\left(\phi_{0}, k\right) - E\left(\psi_{1}, k\right)\right] - \left[2 F\left(\frac{\pi}{2}, k\right) - F\left(\phi_{0}, k\right) - F\left(\psi_{1}, k\right)\right] = 0
\end{equation*}
where $k = \sqrt{(1 + \bar{f})/2}$, and
\begin{align*}
    &\phi = \cos^{-1} \left(\frac{\sqrt{B}}{2 k} \left(\bar{Y} - Y\right)\right), \qquad \phi_{0} = \cos^{-1} \left(\frac{\sqrt{B}}{2 k} \bar{Y}\right)                        && (0 \leq Y \leq \bar{Y})\\
    &\psi = \cos^{-1} \left(\frac{\sqrt{B}}{2 k} \left(Y - \bar{Y}\right)\right), \qquad \psi_{1} = \cos^{-1} \left(\frac{\sqrt{B}}{2 k} \left(2 Y_{0} - \bar{Y}\right)\right) && (\bar{Y} \leq Y \leq 2 Y_{0}) \ .
\end{align*}

\noindent \underline{(2) $2 Y_{0} \leq \bar{Y}$}
\begin{equation*}
    X \left(Y\right) = 
    \begin{cases}
        1 - \dfrac{1}{\sqrt{B}} \left[2 \left\{E\left(\phi_{0}, k\right) - E\left(\phi, k\right)\right\} - \left\{F\left(\phi_{0}, k\right) - F\left(\phi, k\right)\right\}\right]                                                                                                     & \qquad \qquad (\left|\bar{f}\right| \leq 1)\\
        \\
        1 - \dfrac{1}{\sqrt{B}} \left[\dfrac{2}{\tilde{k}} \left\{E\left(\varphi_{0}, \tilde{k}\right) - E\left(\varphi, \tilde{k}\right)\right\} + \dfrac{\tilde{k}^{2} - 2}{\tilde{k}} \left\{F\left(\varphi_{0}, \tilde{k}\right) - F\left(\varphi, \tilde{k}\right)\right\}\right] & \qquad \qquad (\bar{f} > 1)
    \end{cases}
\end{equation*}
The boundary condition of $X \left(Y\right) = 1$ at $Y = 2 Y_{0}$ gives
\begin{align*}
    &2 \left[E\left(\phi_{0}, k\right) - E\left(\phi_{1}, k\right)\right] - \left[F\left(\phi_{0}, k\right) - F\left(\phi_{1}, k\right)\right] = 0                                                                            && (\left|\bar{f}\right| \leq 1)\\
    \\
    &2 \left[E\left(\varphi_{0}, \tilde{k}\right) - E\left(\varphi_{1}, \tilde{k}\right)\right] + \left(\tilde{k}^{2} - 2\right) \left[F\left(\varphi_{0}, \tilde{k}\right) - F\left(\varphi_{1}, \tilde{k}\right)\right] = 0 && (\bar{f} > 1)
\end{align*}
where $\phi$, $\phi_{0}$ and $k$ are same as the above case, and
\begin{align*}
    &\phi_{1} = \cos^{-1} \left(\frac{\sqrt{B}}{2 k} \left(\bar{Y} - 2 Y_{0}\right)\right)                                                                                            && (\left|\bar{f}\right| \leq 1)\\
    &\varphi  = \sin^{-1} \left(\frac{\sqrt{1 - f \left(Y\right)}}{\sqrt{2}}\right), \qquad \varphi_{0} = \frac{\pi - \theta_{d}}{2}, \qquad \tilde{k} = \sqrt{\frac{2}{1 + \bar{f}}} && (\bar{f} > 1) \ .
\end{align*}

\noindent \underline{(3) $\bar{Y} \leq 0$}
\begin{equation*}
    X \left(Y\right) = 
    \begin{cases}
        1 + \dfrac{1}{\sqrt{B}} \left[2 \left\{E\left(\psi_{1}, k\right) - E\left(\psi, k\right)\right\} - \left\{F\left(\psi_{1}, k\right) - F\left(\psi, k\right)\right\}\right]                                                                                         & \qquad \qquad (\left|\bar{f}\right| \leq 1)\\
        \\
        1 + \dfrac{1}{\sqrt{B}} \left[\dfrac{2}{\tilde{k}} \left\{E\left(\psi_{0}, \tilde{k}\right) - E\left(\psi, \tilde{k}\right)\right\} + \dfrac{\tilde{k}^{2} - 2}{\tilde{k}} \left\{F\left(\psi_{0}, \tilde{k}\right) - F\left(\psi, \tilde{k}\right)\right\}\right] & \qquad \qquad (\bar{f} > 1)
    \end{cases}
\end{equation*}
and the boundary condition at $Y = 2 Y_{0}$ gives
\begin{align*}
    &2 \left[E\left(\psi_{0}, k\right) - E\left(\psi_{1}, k\right)\right] - \left[F\left(\psi_{0}, k\right) - F\left(\psi_{1}, k\right)\right] = 0                                                                && (\left|\bar{f}\right| \leq 1)\\
    \\
    &2 \left[E\left(\psi_{0}, \tilde{k}\right) - E\left(\psi_{1}, \tilde{k}\right)\right] + \left(\tilde{k}^{2} - 2\right) \left[F\left(\psi_{0}, \tilde{k}\right) - F\left(\psi_{1}, \tilde{k}\right)\right] = 0 && (\bar{f} > 1)
\end{align*}
where $\psi_{0} = \cos^{-1} \left(- \bar{Y} \sqrt{B} / \left(2 k\right)\right)$.

We can find the contact angles ($\theta_{u}$ and $\theta_{d}$) using the boundary conditions as well as the relation in Eq.(\ref{eqn: area_constraint_with_gravity}) for given $A^{*}$, $B$, and $Y_{0}$. A few profiles of two-dimensional horizontal capillary bridges for $A^{*} = 4$ in the absence and presence of gravity ($B = 0$ and $B = 0.25$) are shown in Fig.\ref{Fig_6}.

\section{Neck, Bulge, and Pinch-off}

\subsection{Without gravity ($B = 0$)}
\begin{figure}[t!]
    \includegraphics[width=1.0\textwidth]{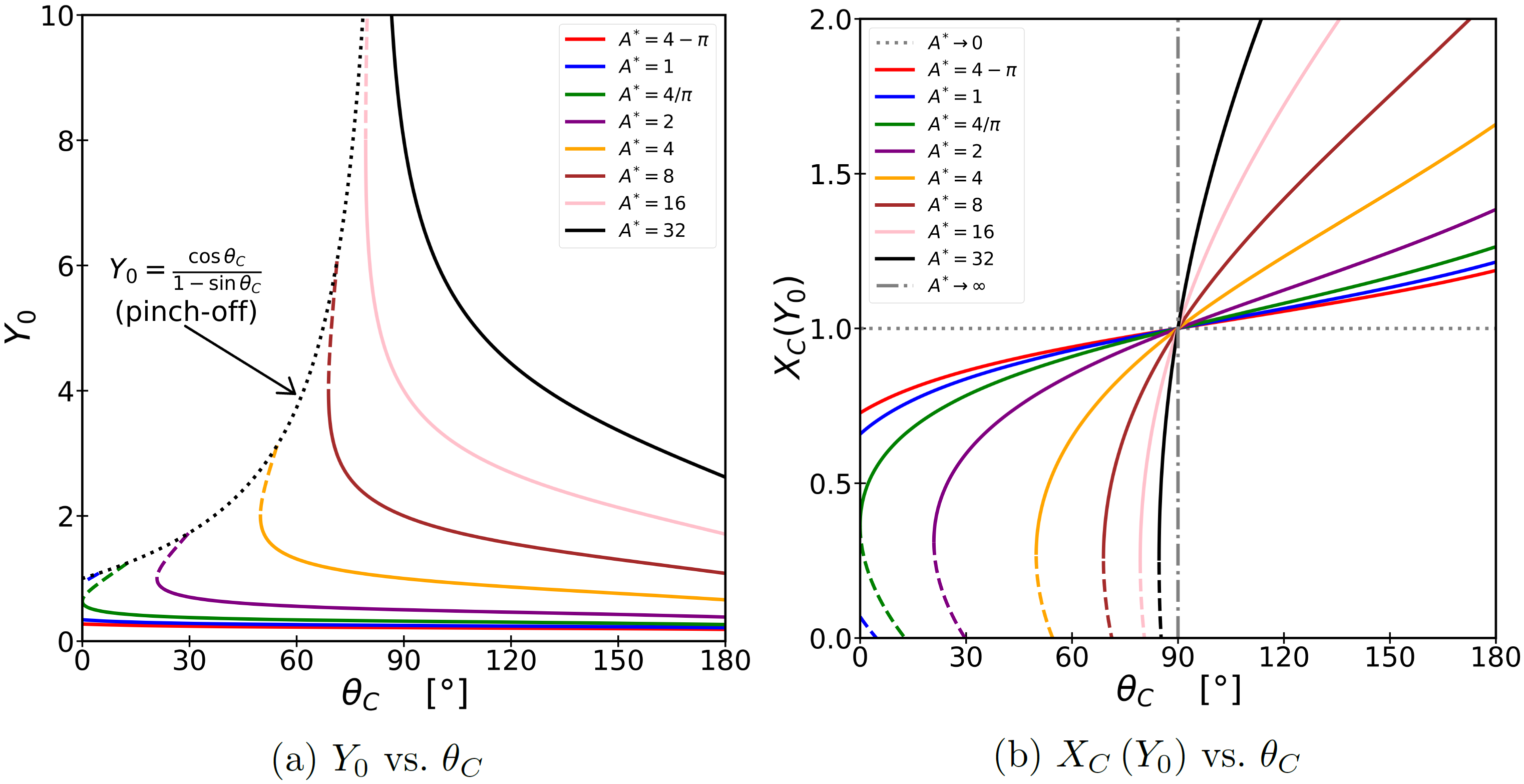}
    \caption{Solid lines are for $X_{C} \left(Y_{0,-}\right)$ and dashed lines are for $X_{C} \left(Y_{0,+}\right)$ where $X_{C} \left(Y_{0, \pm}\right) = 1 - Y_{0, \pm} \left(1 - \sin \theta_{C}\right) / \cos \theta_{C}$.}
\label{Fig_7}
\end{figure}

When there is no gravity, the point at $Y = Y_{0, \pm}$ represents either a neck or a bulge ($X_{C}' \left(Y_{0, \pm}\right) = 0$). 
Depending on the value of $X_{C, \pm} \equiv X_{C} \left(Y_{0, \pm}\right)$ compared to the position of the contact point ($X=1$) (or equivalently, depending on the sign of $X''_{C} \left(Y_{0, \pm}\right)$), we can divide the general shapes of the bridges into three separate cases: \\
(1) If $0 \leq X_{C, \pm} < 1$ (i.e. $0^{\circ} \leq \theta_{C} < 90^{\circ}$), the profile has a neck. A pinch-off ($X_{C, +} = 0$) can occur only in this case. \\
(2) If $X_{C, \pm} = 1$ (i.e. $\theta_{C} = 90^{\circ}$), the profile is rectangular. \\
(3) If $X_{C, \pm} > 1$ (i.e. $90^{\circ} < \theta_{C} \leq 180^{\circ}$), the profile has a bulge. \\
The positions of the neck or the bulge, $X_{C, \pm}$ and $Y_{0, \pm}$, are shown in Fig.\ref{Fig_7} as a function of $\theta_{C}$.

In Fig.\ref{Fig_7} (a), only $Y_{0,-}$ exists for all $\theta_{C}$ when $0 \leq A^{*} \leq 4 - \pi$. For $4 - \pi < A^{*} < 4 / \pi$, there is an interval of $\theta_{C}$ ($0^{\circ} \leq \theta_{C} \leq \theta_{C, \text{pinch}}$) where both $Y_{0, +}$ and $Y_{0, -}$ exist for a single $\theta_{C}$. $Y_{0,-}$ exists in $0^{\circ} \leq \theta_{C} \leq 180^{\circ}$, but $Y_{0,+}$ exists only for an interval of $\theta_{C}$ ($0^{\circ} \leq \theta_{C} \leq \theta_{C, \text{pinch}}$). So the two curves of $Y_{0,-}$ and $Y_{0,+}$ are not continuous. For $A^{*} \geq 4 / \pi$, there is a point where $\partial \theta_{C} / \partial Y_{0} = 0$, so the two curves, $Y_{0, \pm}$, become continuous. In this case, $Y_{0,-}$ and $Y_{0,+}$ exist only when $\theta_{C, \text{min}} \leq \theta_{C} \leq 180^{\circ}$ and $\theta_{C, \text{min}} \leq \theta_{C} \leq \theta_{C, \text{pinch}}$, respectively. 

In Fig.\ref{Fig_7} (b), we can see that pinch-off occurs only for $Y_{0,+}$ at $\theta_{C} = \theta_{C, \text{pinch}}$, and in addition a pinch-off does not occur when $0 \leq A^{*} \leq 4 - \pi$. A neck appears for both of $Y_{0, \pm}$, but a bulge can only appear for $Y_{0,-}$.

\subsection{With gravity ($B > 0$)}
The positions of a neck or a bulge $\left(X_{1} \equiv X \left(Y_{1}\right), Y_{1}\right)$ are defined by the relation $X' \left(Y_{1}\right) = 0$. So, we have 
\begin{equation} \label{eqn: Y_1_pm}
    Y_{1, \pm} = \bar{Y} \pm \sqrt{\bar{Y}^{2} + \frac{2}{B} \cos \theta_{d}}
\end{equation}
where $\bar{Y}$ is defined in Eq.(\ref{eqn: f(Y)}). A neck ($X_{1} < 1$) or a bulge ($X_{1} > 1$) can appear only when its position $Y_1$ is in the interval $0 \leq Y_{1, \pm} \leq 2 Y_{0}$. When there is gravity, the rectangular solution ($X_{1} = 1$) is not allowed since $\theta_{u} < \theta_{C} < \theta_{d}$ due to contact angle hysteresis. The sign of the second derivative of the profile function $X'' \left(Y_{1, \pm}\right) = f' \left(Y_{1, \pm}\right) \left[1 - \left(f\left(Y_{1, \pm}\right)\right)^{2}\right]^{-3/2}$ is equal to that of $f' \left(Y_{1, \pm}\right)$. And combining Eq.(\ref{eqn: f(Y)}) and Eq.(\ref{eqn: Y_1_pm}), we have
\begin{equation*}
    f' \left(Y_{1, \pm}\right) = B \left(Y_{1, \pm} - \bar{Y}\right) = \pm \sqrt{2 B \bar{f}}
\end{equation*}
where $\bar{f}$ is given in Eq.(\ref{eqn: f(Y)}). A capillary bridge under gravity should have at least a neck or a bulge, or both, the right hand side of the above equation should always be real, implying $\bar{f} \geq 0$. Thus, a neck and a pinch-off can occur for $Y_{1, +}$ while a bulge does for $Y_{1, -}$.

\nocite{*}

\bibliography{DeterminationYoungAngle.bib}

\end{document}